# Long-lived optical phonons in ZnO studied with impulsive stimulated Raman scattering

C. Aku-Leh, J. Zhao, and R. Merlin

*FOCUS Center and Department of Physics, The University of Michigan,
Ann Arbor, MI 48109-1120, USA*

J. Menéndez

*Department of Physics and Astronomy, Arizona State University, Tempe, AZ, 85287-1504, USA*

M. Cardona

*Max-Planck-Institut für Festkörperforschung, Heisenbergstraße 1, 70569 Stuttgart, Germany*

The anharmonic properties of the low-frequency $E_2$ phonon in ZnO were measured using impulsive stimulated Raman scattering. At 5 K, the frequency and lifetime are $(2.9787 \pm 0.0002)$ THz and $(211 \pm 7)$ ps. The unusually long lifetime and the high accuracy in the determination of the frequency hold promise for applications in metrology, quantum computation and materials characterization. The temperature dependence of the lifetime is determined by two-phonon up-conversion decay contributions, which vanish at zero temperature. Results suggest that the lifetime is limited by isotopic disorder and that values in the nanosecond range may be achievable in isotopically-pure samples.

The development of commercial laser sources with femtosecond pulses has made it possible to generate and control coherent excitations in semiconductors [1,2,3]. This capability creates exciting opportunities for applications in several fields and provides a powerful tool for the study of fundamental interactions. In particular, detailed schemes have been proposed to take advantage of coherent fields for ultrafast optical-data processing which require long-lived excitations for information storage and vanishing group velocities for miniaturization [4]. Recent impulsive stimulated Raman scattering (ISRS) experiments in wurtzite ZnO [5] and GaN [6] suggest that their doubly-degenerate low-frequency (LF) $E_2$ phonons have exactly these properties. In ZnO, Lee and coworkers [5] find a lifetime of $\tau = 29.2$ ps at room temperature, about an order of magnitude longer than typical lifetimes for Raman phonons in semiconductors [7]. Similarly, the corresponding $E_2$ mode in GaN was found to have a lifetime of $\tau = 70$ ps [6]. Long-lived optical phonons are potentially of interest for applications in quantum information science. In particular, doubly and triply-degenerate modes could play a role similar to that of photons in quantum cryptography [8] and, like photons in cavity-quantum-electrodynamics schemes for quantum computing [9], phonons could also be drawn to mediate the interaction between qubits. Within this context, we note that phonons associated with the center-of-mass motion of a system of trapped ions have been utilized to mediate their interaction and produce many-particle entangled states [10,11].

In this paper, we report an ISRS study of the temperature dependence of the frequency and lifetime of the lowest-lying $E_2$ phonon in ZnO, referred to in the following as $E_2$(LF). We find a dramatic lifetime increase as the temperature is decreased, up to a value of $\tau = 211$ ps at 5 K. This strong temperature dependence is explained in terms of an anharmonic decay mechanism dominated by phonon up-conversion processes. The up-conversion contribution to the phonon lifetime is expected to vanish as the temperature approaches zero, and indeed our measured low-temperature lifetime is close to the predicted lifetime due to isotopic disorder in natural ZnO. Hence we expect isotopically pure ZnO crystals to have even longer phonon lifetimes, probably in the nanosecond range. Our study also reveals that the frequency of long-lived phonons detected with ISRS can be determined with a precision of a few parts per million. On that account ISRS of long-lived phonons may have additional applications in metrology (although the quality factor for ZnO is ~ 2000, i. e., 1-2 orders of magnitude below those for quartz oscillators [12] which, however, operate at much lower frequencies), as well as for characterizing the purity and perfection of semiconductor crystals.

Measurements were performed on a $10\times5\times1$ mm$^3$ ZnO single crystal, with the *c*-axis parallel to the larger side of the parallelepiped, in the temperature range 5-300 K using a continuous-flow liquid-He optical cryostat. Pressures inside the dewar were kept constant at ~28 kPa. The laser source was a regenerative amplifier seeded by a Ti:Sapphire oscillator which provided ~ 70 fs pulses of central wavelength 803 nm at the repetition rate of 250 kHz. (note that the bandwidth of our pulses is a factor of ~ 5 larger than the $E_2$(LF) frequency, but too small to observe oscillations associated with the high-frequency $E_2$ phonon at ~ 13.3 THz [5]). Since the laser energy is well below the band gap of ZnO, the generation mechanism of coherent phonons is ISRS [2]. Data were obtained using a conventional pump-probe setup. The pump pulse creates coherent vibrations, which modulate temporally and spatially the refractive index of the material. In turn, this modulation leads to a change in the transmission of the probe beam, $\delta T(\omega)$, which we measured as a function of the light frequency $\omega$ and the delay between the two pulses, using standard lock-in methods. Because the photon number is conserved in Raman processes and, hence, $\int \delta T(\omega)d\omega = 0$ [2], we used a balanced detector to determine the difference between the two half portions of the probe spectrum; this difference is called Δ$T$. To satisfy the Raman selection rules for $E_2$ phonons, the pump and probe polarization were both parallel to the $a$-axis. Light penetrated the largest face of the crystal through a circular spot of diameter 66 μm (pump) and 38 μm (probe), using an average power of 10 mW (pump) and 2.5 mW (probe).

In Fig. 1 we show an example of coherent phonon oscillations associated with $E_2$(LF). For a strictly impulsive force, δ($t$), a mode of frequency $f_0$ behaves as $Q \propto e^{-\Gamma t} \sin(2\pi f_0 t)$, where $Q$ is the phonon coordinate and $\Gamma = 1/\tau$ and, since $\Delta T \propto dQ/dt$ [2], we get $\Delta T \propto e^{-\Gamma t} \cos(2\pi f_0 t)$. Thus, it is in principle possible to relate the *real* part of the Fourier transform of the transmitted signal to the spontaneous Raman lineshape, determined by the imaginary part of the phonon self-energy [2,13]. However, uncertainty in the determination of time zero and artifacts produced by multiple reflections of the pump beam cause the Fourier transform components to become a mixture of the real and imaginary parts of the self-energy. This problem can be partially circumvented by computing the power spectrum, but such approach is not entirely satisfactory in anharmonicity studies because the width of the power spectrum is not directly related to the width of the imaginary part of the self-energy. To solve this problem, we developed a phase-correction algorithm similar to the one used for the analysis of nuclear-magnetic-resonance (NMR) data [14]. This algorithm yields a corrected Fourier spectrum that can be directly compared with theoretical calculations of the Raman lineshape. In Fig. 2 we show the corrected Fourier transform of the coherent phonon oscillations, which we denote as the Raman lineshape. In addition to applying the above-mentioned phase correction, we have optimized the density of points in Fig. 2 by using a standard NMR-FFT (fast Fourier transform) zero-filling approach [9]. We fit the resulting lineshape with a Lorentzian profile of the form $A\left[(f-f_0)^2 + (\Gamma/2\pi)^2\right]^{-1}$, from which we obtain the mode frequency $f_0$ and the full width at half maximum FWHM = $\Gamma/\pi$ of the Raman lineshape. These parameters are plotted in Fig. 3 as a function of temperature. Notice the high precision of the data. For $T = 5$ K, for example, the mode frequency is $f_0 =$ (2.9789 ± 0.0002) THz [(99.390 ± 0.007) cm$^{-1}$] and FWHM = (1.602 ± 0.052)×10$^{-3}$ THz [(0.0535 ± 0.002) cm$^{-1}$]. Thus the frequency is obtained with five significant figures. The largest source of error in the frequency is the uncertainty in the length of the delay line, followed by corrections arising from fluctuations in the refractive index of air. Since there are relatively simple ways to reduce the experimental uncertainty, the ultimate precision of the ISRS technique has not been reached. The ability to measure phonon frequencies with very high precision may open up new applications for phonon spectroscopy in materials.

In order to validate our Fourier-transform procedure, we also obtained the lifetimes directly from the time-domain data by fitting 4-ps-segments of data (see inset of Fig. 1) with an expression of the form $A \cos(2\pi f_0 t + \alpha)$ and plotting the resulting average amplitude $A$ versus the average delay time $t_c$ (the time at the center of the interval), as shown in Fig. 4. The $A(t_c)$ curve is then fitted with a single exponential $B \exp(-\Gamma t_c)$, from which $\Gamma$ can be extracted. The results are plotted as squares in Fig. 3a. As expected, the broadening parameters obtained from the time domain analysis are nearly the same as those obtained from the frequency domain fits. Notice that the value of the lifetime at room temperature, $\tau = \Gamma^{-1} = 34.1$ ps, is somewhat longer than 29.2 ps, as reported in [5].

The temperature dependence of the FWHM ($\equiv \Gamma/\pi$) of the Raman lineshape is usually fitted using the following expression for $\Gamma$ [15,16]:

$$\Gamma(T) = \Gamma_0 + \Gamma_{DN}\left(1 + n_1 + n_2\right) \quad (1)$$

where $n_{1,2} = \left[\exp(hf_{1,2}/k_B T)-1\right]^{-1}$ is the phonon occupation number, or Bose factor, and the frequencies satisfy $f_1 + f_2 = f_0$. The term $\Gamma_0$ accounts for a temperature-independent broadening caused by defects, including isotopic mixture. The second term in the right hand side models a phonon *down-conversion* process induced by third-order anharmonicity. It represents the decay of the optical mode into two lower-frequency phonons. The down-conversion term is actually an appropriately weighted sum over all allowed decay channels [17,18] satisfying conservation of energy and crystal momentum. Equation (1) replaces the sum over individual channels by a single effective decay channel. Recent *ab initio* predictions [18,19,20] (which have revolutionized the field of anharmonicity studies) provide a strong justification for this approach, since they show indeed that the decay frequencies cluster around a few pairs (often only one pair) of values. The phonon occupation numbers vanish at low temperatures, and the down-conversion contribution approaches a constant value $\Gamma_{DN}$. The magnitude of $\Gamma_{DN}$ is determined by an anharmonic matrix element and the density of two-phonon states at the frequency $f_0$ of the decaying phonon. When this frequency is sufficiently low, the only possible decay products are combinations of acoustic phonons, whose density of states is a rapidly decreasing function of frequency. The anharmonic matrix elements also vanish in the low-frequency limit [20], so that the lifetime due to down-conversion processes should increase sharply as the frequency $f_0$ decreases.

In ZnO, the $E_2$(LF) frequency $f_0$ is much lower than the maximum optical phonon frequency of ~ 17.6 THz for the $A_1$ longitudinal-optical mode [21]. Therefore, we expect phonon down-conversion processes to make a small contribution to the phonon lifetime. On the other hand, the existence of phonon branches with frequencies *higher* than $f_0$ enables an additional third-order anharmonic process, mainly the decay of the Raman phonon with the simultaneous annihilation of a phonon of frequency $f_3$ and creation of a phonon with frequency $f_4$. By including these *up-conversion* processes, the expression for $\Gamma$ becomes

$$\Gamma(T) = \Gamma_0 + \Gamma_{DN}\left(1 + n_1 + n_2\right) + \Gamma_{UP}\left(n_3 - n_4\right) \qquad (2)$$

Notice that the up-conversion term vanishes at zero temperature, so that even with the inclusion of this term the low-temperature lifetime in defect-free materials should be dominated by down-conversion processes. To the best of our knowledge, the up-conversion contribution to the anharmonic decay of Raman-active phonons in semiconductors has never been conclusively detected. In particular, the Raman-active longitudinal-optical phonons in most cubic semiconductors have the highest vibrational frequency in the crystal, and therefore the up-conversion process is forbidden for these phonons. Predominance of the up-conversion channel has been recently established in measurements of the lifetime of the TA(*X*) phonons in Ge and confirmed by *ab initio* calculations [22].

Given the different behaviors for $T \rightarrow 0$, a fit with Eq. (2) can easily establish the relative weight of the down- and up-conversion channels if the defect contribution can be neglected. Unfortunately, our ZnO sample possesses a natural distribution of Zn and O isotopes, so that the first term in Eq. (2) cannot be neglected. A fit with the full Eq. (2) expression is difficult and leads to large errors in the fitting parameters. We therefore adopt the following approach. First, we fit the temperature dependence of the inverse lifetime by neglecting the up-conversion contribution. For this, we assume $f_1 = f_2 = f_0/2$ (the so-called Klemens Ansatz [23]) and we obtain the dotted line in Fig. 3a, with $\Gamma_0/\pi = 0.0006$ THz and $\Gamma_{DN}/\pi = 0.0009$ THz. (A fit with $f_1 = xf_0$, $f_2 = (1-x)f_0$, with $x$ an adjustable parameter as in [24], yields $x = 0.7$, but the curve is hardly distinguishable from the dotted line in Fig. 3a). Next, we neglect the down-conversion channel and fit the experimental results with an expression that contains only the first and third terms in Eq. (2). From inspection of the calculated phonon density of states (DOS) for ZnO [21], we identify two possible sets of frequencies for the up-conversion process: $f_3 = 13.4$ THz; $f_4 = 16.4$ THz and $f_3 = 4.47$ THz; $f_4 = 7.45$ THz. The first combination gives a poor fit, but the second one yields the solid line in Fig. 3a, with parameters $\Gamma_0/\pi = 0.0015$ THz and $\Gamma_{UP}/\pi = 0.015$ THz. It is apparent from a comparison of the two curves in Fig. 3a that the up-conversion fit is in better agreement with experiment. This strongly suggest that phonon up-conversion plays a dominant role in the anharmonic decay of the $E_2$(LF) phonons in wurtzite materials. Our findings are consistent with the work on TA(*X*) phonons in Ge mentioned above, for which up-conversion process were found to dominate and the low-temperature lifetime estimated at 2 ns [22]. The $E_2$(LF) phonons in the wurtzite structure are, in some sense, equivalent to TA(*L*) phonons in the zincblende structure. Since the kinematic decay restrictions for TA(*L*) and TA(*X*) are similar, the up-conversion channel should also be dominant for TA(*L*) phonons in zincblende as well as for $E_2$(LF) phonons in wurtzite.

Theoretical estimates of the isotopic contribution to the linewidth of the $E_2$(LF) phonon provide additional support for the dominance of up-conversion processes. The corresponding phonon eigenvector consists mainly of Zn-displacements, so it is the isotopic disorder in the Zn sublattice which affects the phonon lifetime. Our natural ZnO crystals contain 48.6% $^{64}$Zn, 27.9% $^{66}$Zn, 4.1% $^{67}$Zn, 18.75% $^{68}$Zn, and 0.6% $^{70}$Zn [25], and therefore the isotopic contribution to $\Gamma$ should be sizable. A recent calculation by Serrano and co-workers [21], using *ab initio* phonon density of states, yields $\Gamma_0/\pi = 0.0018$ THz. This is very close to the experimental value for the FWHM at low temperatures, suggesting that at these temperatures the intrinsic anharmonic linewidth is significantly small, as expected from an up-conversion-dominated decay mechanism. This exciting result indicates that in isotopically pure samples the lifetime of the $E_2$(LF) phonon could be an order of magnitude longer (and perhaps even longer) than in our natural crystals. Lifetimes in the nanosecond range have not yet been reported for optical phonons in semiconductors.

In summary, we have shown that the $E_2$(LF) phonon lifetime in ZnO is longer than 200 ps at low temperatures, and probably in the ns range in isotopically pure crystals. Long low-temperature $E_2$(LF) lifetimes should also be expected in other wurtzite materials, including CdS, CdSe, GaN and AlN. Moreover, more complex tetrahedral semiconductors (such as the chalcopyrite materials of the I-III-VI$_2$ family) display a number of optical phonon branches with even lower frequencies [26]. Hence these materials are also strong candidates for coherently driven long-lived optical phonons. Our results demonstrate the extraordinary potential of the ISRS technique, not only for anharmonicity studies, but also for new spectroscopic applications. Phonon frequencies determined with the precision than can be attained with ISRS should be sensitive to extremely small crystalline perturbations that hitherto were considered to be beyond the reach of phonon spectroscopy.

This work was supported by the AFOSR under contract F49620-00-1-0328 through the MURI program and by the NSF Focus Physics Frontier Center. J. M. would like to acknowledge support through the NSF FOCUS Fellows program at the University of Michigan.

# FIGURE CAPTIONS

FIG. 1 (color online). Differential change in transmission at 20 K and 292 K showing coherent $E_2$(LF) oscillations in ZnO. The black rectangle indicates the range of the scan shown in the inset.

FIG. 2 . Raman lineshapes obtained from the time-domain coherent phonon oscillations (circles). The solid lines represent Lorentzian fits to the data. An arbitrary vertical offset was applied to the data for clarity.

FIG. 3. a) Circles: Full width at half maximum (FWHM) of the Raman lineshapes obtained from Lorentzian fits to the frequency domain data. Notice that the vertical errors are smaller than the diameter of the circles. Squares: FWHM obtained from time-domain fits to the coherent oscillation amplitude using a simple decaying exponential. See text for fit details and also Fig. 4.

FIG. 4. Average oscillating amplitude over 4 ps segments as a function of the average delay time for the segment. Solid line represents the best fit with a single decaying exponential.

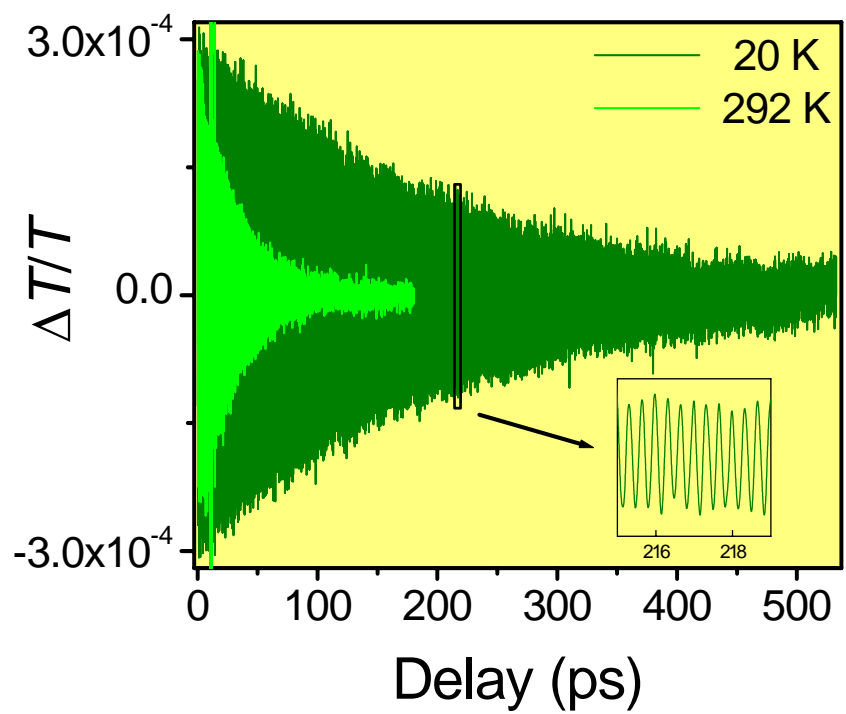

**Figure 1** Aku-Leh et al.

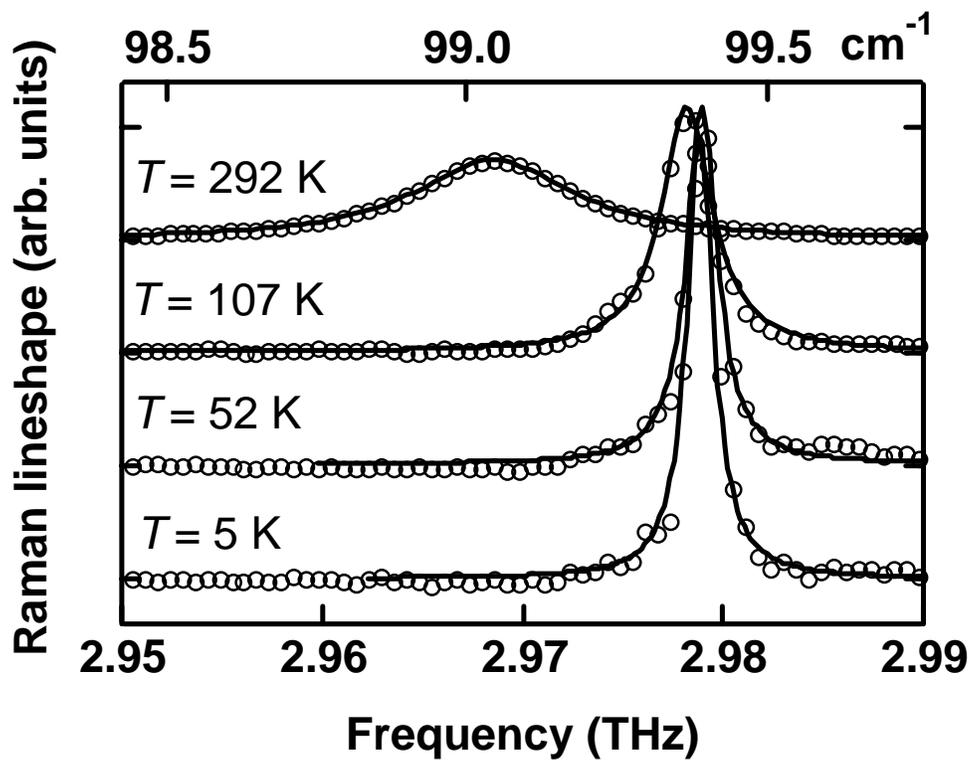

**Figure 2** Aku-Leh et al.

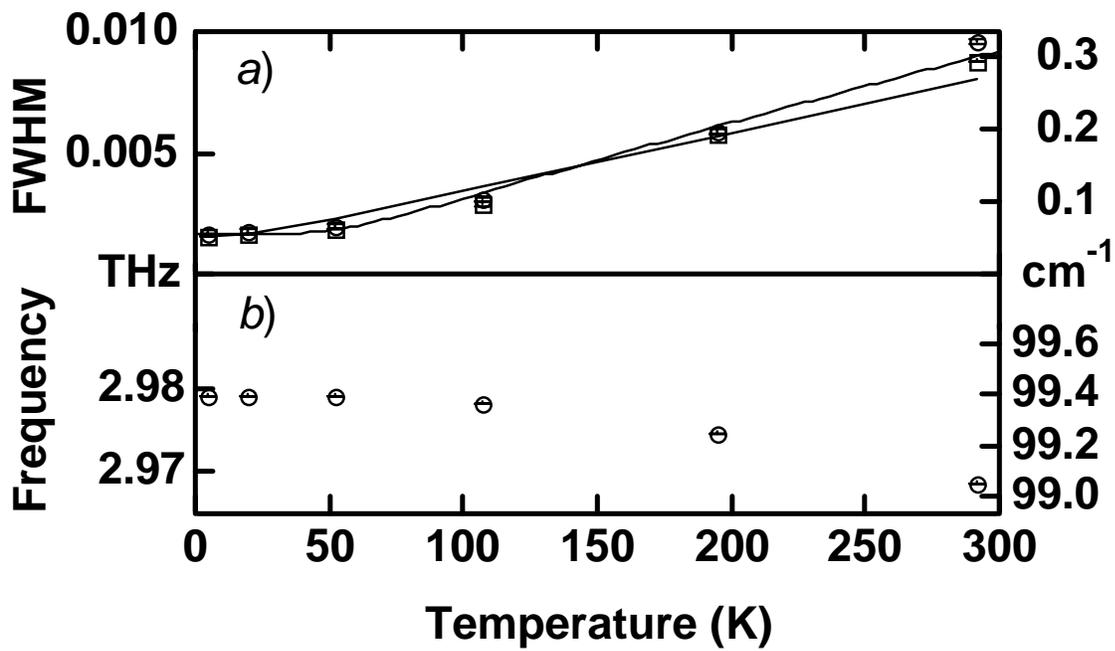

**Figure 3** Aku-Leh et al.

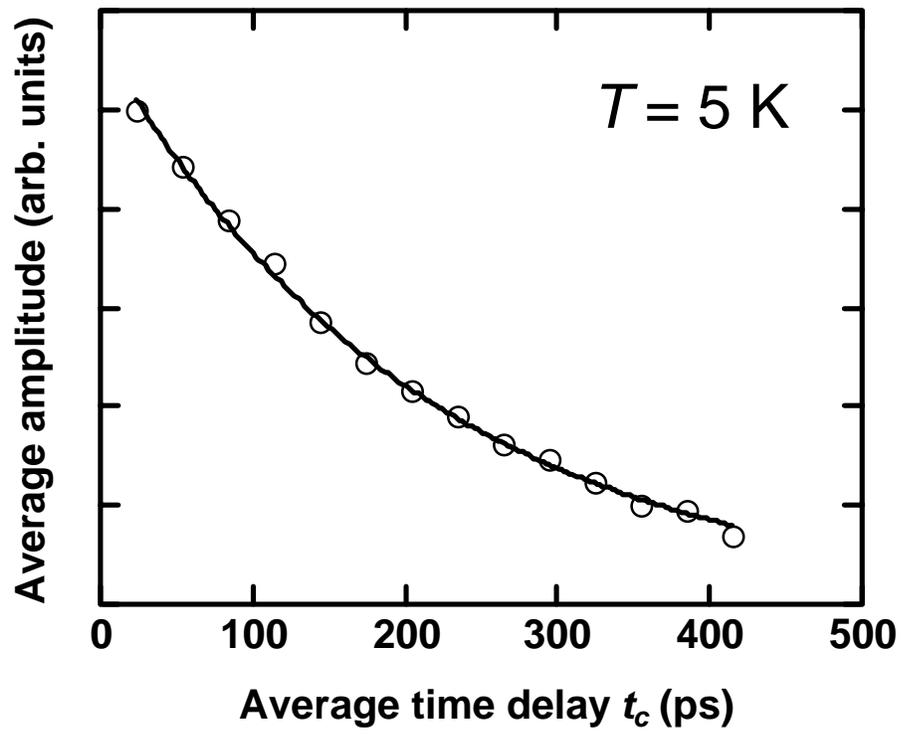

**Figure 4** Aku-Leh et al.